\documentclass[acmsmall]{acmart} %

\AtBeginDocument{%
  \providecommand\BibTeX{{%
    \normalfont B\kern-0.5em{\scshape i\kern-0.25em b}\kern-0.8em\TeX}}}

\newcounter{rq}[section]
\newenvironment{rqenv}[1][]{\refstepcounter{rq}
   \textbf{RQ\therq: #1}}{}

\usepackage{wrapfig}

\begin{document}

\title{Engaging Teachers to Co-Design Integrated AI Curriculum for K-12 Classrooms}

%
%
%
%
%
%

\author{Jessica Van Brummelen}
\authornote{Both authors contributed equally to this research.}
\affiliation{%
  \institution{Massachusetts Institute of Technology}
  \streetaddress{77 Massachusetts Avenue}
  \city{Cambridge}
  \state{MA}
  \country{USA}}
\email{jess@csail.mit.edu}

\author{Phoebe Lin}
\authornotemark[1]
\affiliation{%
  \institution{Harvard University}
  \streetaddress{13 Appian Way}
  \city{Cambridge}
  \state{MA}
  \country{USA}}
\email{phoebelin@gsd.harvard.edu}


%
%
%
%
%
%

%
%
%
\begin{abstract}
Artificial Intelligence (AI) education is an increasingly popular topic area for K-12 teachers. However, little research has investigated how AI education can be designed to be more accessible to all learners. We organized co-design workshops with 15 K-12 teachers to identify opportunities to integrate AI education into core curriculum to leverage learners' interests. During the co-design workshops, teachers and researchers co-created lesson plans where AI concepts were embedded into various core subjects. We found that K-12 teachers need additional scaffolding in the curriculum to facilitate ethics and data discussions, and value supports for learner engagement, collaboration, and reflection. We identify opportunities for researchers and teachers to collaborate to make AI education more accessible, and present an exemplar lesson plan that shows entry points for teaching AI in non-computing subjects. We also reflect on co-designing with K-12 teachers in a remote setting.

\end{abstract}



\begin{teaserfigure}
  \includegraphics[width=\textwidth]{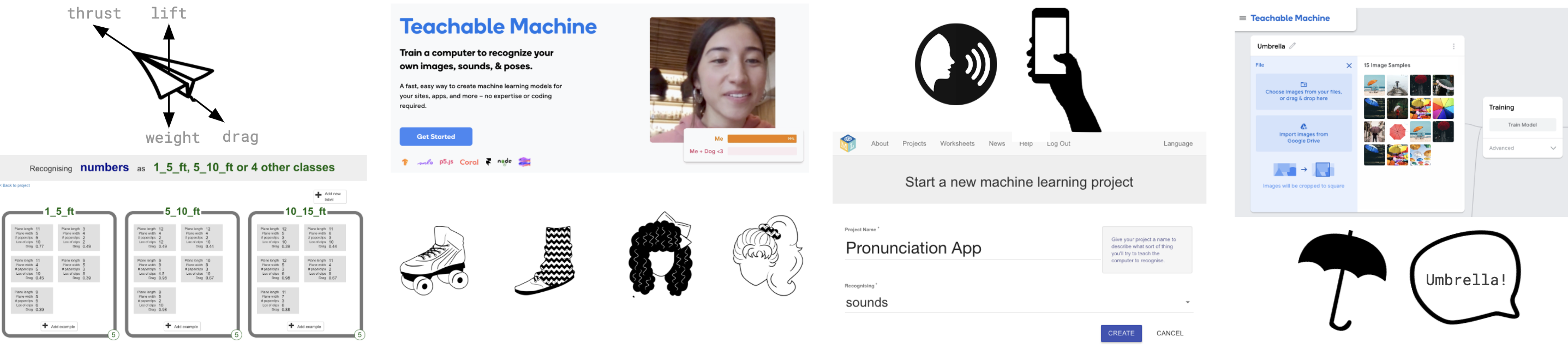}
  \caption{Representations of the four integrated curricula \cite{gist-appendix,ml4kids,teachable-machine}. %
  \emph{Left}: The ``exemplar'' physics and AI curriculum. \emph{Mid-left}: Social studies and AI curriculum. \emph{Mid-right}: ESL and AI curriculum. \emph{Right}: Literacy and AI curriculum for students with learning disabilities.}
  \Description{Images of AI tool websites, including Machine Learning for kids \citep{ml4kids} and Teachable Machine \citep{teachable-machine} alongside a paper airplane, outfits, a voice application, and an umbrella.}
  \label{fig:teaser}
\end{teaserfigure}

\maketitle

\section{Introduction}
Artificial intelligence (AI) education is becoming an increasingly popular subject in the eyes of educators due to the rapid integration of AI technologies in user-facing services and products \cite{ai-edu-world-goel,k-12-ai-edu-touretzky,accelerated-ai-edu-yang}. Researchers have called for formal K-12 education to prioritize AI literacy and teach children to interact with AI using a critical lens \cite{zimmerman}. The AI4K12 research community has also published guidelines for what AI concepts K-12 curriculum should cover, known as the \emph{Big AI Ideas}, and calls for AI researchers to help teachers and students understand AI \cite{touretzky2019envisioning}. As children interact more with AI technologies, it is critical that they are able to recognize AI, understand how AI algorithms work, use those algorithms to solve problems meaningful to them, and evaluate the impact of AI technologies on society \cite{ali2019constructionism}.

Teachers of all subjects should feel empowered to teach AI curriculum, yet teachers often feel they lack sufficient understanding to teach AI or the capacity to include more curriculum on top of their existing curriculum \cite{teacher-education}. Despite the proliferation of tools and AI curriculum in response to the recent calls to action, few are widely implemented due to challenges in the classroom that prevent these curricula from being accessible \cite{ed-innovation-codesign}. In order to introduce new practices, researchers and developers should consider the contexts of teachers and invest in additional supports to facilitate the accessibility of AI resources for teachers.

Similarly, AI as a discipline can span many other topics, such as government, journalism, and art \cite{gan-art-hertzmann,ai-govt-chen,journalism-ai-stray}, therefore AI should not be confined to just computing subjects such as computer science or data science. Tools and curriculum today often teach AI as an extension of computer science curricula or as standalone curricula that is difficult to adjust to other contexts \cite{pic-danny,personalizing-homemade-bots,irobot}. Adapting those tools and curriculum then becomes especially difficult for teachers who teach core subjects, including English, math, social studies, and science, and may not have any AI experience. The lack of integrated AI curricula in core subjects has become one of the barriers to exposing AI to students with little access to computing disciplines. 

In this paper, we partner with K-12 teachers to design AI curriculum that is integrated with core subjects. We aim to empower all teachers to incorporate AI into their classrooms and leverage learners' interests for other subjects through a two-day co-design workshop with 15 teachers from different schools. We set out to understand what is necessary and valuable to K-12 teachers to effectively implement integrated AI curricula, and co-create lesson plans that address those needs and values. Specifically, our research questions are:

\begin{rqenv}\label{RQ:teaching-needs}
    How might we address K-12 teachers' values and considerations when designing AI curriculum? (Teaching needs)
\end{rqenv}

\begin{rqenv}\label{RQ:integrated-curriculum}
    How might AI be integrated into core subject curriculum? (Integrated curriculum design)
\end{rqenv}

To answer these research questions, we organized a multi-session workshop that spanned two days with fifteen teachers who teach various subjects. The first day of the workshop involved presentations and group discussions to level set everyone's basic understanding of AI. Between the first and second day of the workshop, participants were asked to complete a brainstorming assignment where they identified curriculum of their own to use as a potential base for an integrated AI curriculum. During the second day of the workshop, we split participants into three small groups to work together and design a lesson plan that integrates AI into a non-computing subject curriculum. The co-design process revealed when teachers design curriculum, they consider four practical needs: evaluation, engagement, logistics, and collaboration. Furthermore, our analysis of the co-designed lesson plans showed opportunities for connections between AI and a core subject, with three points of integration: data, reflection, and scaffolding for ethics. 

The contributions of this work are (1) identifying the values and needs of K-12 teachers teaching AI in the classroom and opportunities to address them, (2) showing an exemplar integrated AI curriculum as an output of the co-design session \cite{gist-appendix}, and (3) reflecting on co-design sessions involving K-12 teachers in a remote setting to solicit design considerations of AI curriculum.

\section{Related Work}
To the authors' knowledge, there are no papers describing a co-design process with teachers to integrate AI concepts into core curriculum, and few papers that purposefully integrate AI concepts into core curriculum. Other related research includes the development of AI education tools and curricula, as well as co-design of other course materials with teachers.

\subsection{AI Education for K-12}
Many K-12 AI tools and curricula exist as standalone products or extend computer science curriculum. 
Two widely-used AI teaching tools are \emph{Teachable Machine} \cite{teachable-machine} and \emph{Machine Learning for Kids} \cite{ml4kids}, which empower learners to develop classification models without needing to program. Other standalone AI teaching tools include \emph{Any-Cubes}, which are toys to teach machine learning (ML) concepts \cite{any-cubes-toy-scheidt}; \emph{Calypso for Cozmo}, which is AI curriculum for a toy robot \citep{cozmo}; and extensions for \emph{MIT App Inventor}, which enable students to develop AI-powered mobile apps \cite{ai-appinv}. Each of these tools could be integrated and taught in core classes; however, are presented as standalone AI tools.

In terms of K-12 AI education research involving instructors, most involve researchers rather than K-12 teachers as the instructors, and likely miss valuable expertise and feedback from professionals who have worked in the classroom.
Nevertheless, some works involving K-12 teachers include an AI summer program for high school girls \citep{gender-diversity-ai-summer-vachovsky}, an AI engineering course for high school students \citep{sperling2012integrating}, and a STEM workshop for middle school students \citep{kids-making-ai}. Each of these studies saw value in engaging with K-12 teachers. 

Other works have also integrated %
core curriculum content into AI tools and curricula; however, most of these involve researchers as instructors and are often not in regular classroom settings. For example, one physical education curriculum involves students developing sports gesture classification models with researchers as facilitators \citep{sports-ml-zimmermann}. Another science-based curriculum involves students teaching a conversational agent about animals, and observing it classify the animals into ecosystems with researchers as facilitators \citep{zhorai}. %
Although these works are state-of-the-art in K-12 AI education, it is unknown whether they are suitable for K-12 classrooms, since they have not been tested in regular classrooms and teachers were not involved in the design process. %

In our literature review, we found one example of AI curriculum that was both integrated into a core course and designed or taught by K-12 teachers alongside researchers. This curriculum involved AI and science concepts, and was taught in Australian K-6 classrooms \citep{multiyear-k-6-heinze}. Although this example is insightful, much further research is needed to integrate teacher expertise and address widespread, integrated AI curriculum in K-12 classrooms \citep{teacher-education}.

\subsection{Co-Design in Education}
Although K-12 AI education has not yet benefited from tools and curriculum co-designed with K-12 teachers, other areas of education have. For example, in one study, researchers collaborated with teachers to develop new science curriculum materials. Researchers recognized the value in teachers' K-12 expertise and in promoting their agency throughout the design process \citep{science-codesign-severance}. Another science curriculum co-design study argued that the process of working with teachers had substantial effects on adoption of the tools and curricula, in addition to bringing social value and innovative ideas \citep{co-design-science-durall}. In order to catalyze such benefits, one paper presents key considerations to co-designing with teachers. These include addressing a ``concrete, tangible innovation challenge'', investigating ``current practice and classroom contexts'', and involving a ``central accountability for the quality of the products of the co-design'', among others \citep{ed-innovation-codesign}. In our study, we utilize these considerations and present a co-design for AI-integrated core curricula development.

\section{Method}
We conducted a two-day co-design workshop with fifteen instructors, ranging from K-12 teachers to educational directors. Participants completed pre-work before each day's activities, as well as pre- and post-workshop surveys. The co-design activity was split into three smaller group sessions to enable us to better identify differences in value and process of different teachers. This study was approved by the university’s Institutional Review Board (IRB).

\subsection{Participants}
Fifteen teachers participated in the study, whom we recruited from a mailing list and our personal network. The only inclusion criteria was that they teach or previously taught in a K-12 classroom and were able to commit to the time and pre-work for the two-day workshop. Seven participants identified as female, four participants identified as male, and the remaining did not say. Their age ranged from 25 to 50 (M = 40.6, SD = 11.6). In our selection process, we prioritized participants who primarily taught non-computing subjects, such as English language arts (ELA), and then participants who taught computer science, with the idea that small group sessions could have diverse perspectives. Their work background %
is detailed in Tab. \ref{tab:participants}. All participants provided informed consent to participate in compliance with our institution’s IRB. As the workshop was conducted in groups, we collected participants' availability and selected times where the greatest number of participants could join. Each day of the workshop lasted 2.5 hours. Every workshop session involved two researchers.

\begin{table}[htb!]
\small
\caption{Participants were selected to represent diverse profiles and/or subject areas.}\label{tab:participants}
\begin{tabular}{l|l l l l}
{\textbf{ID}} & {\textbf{Grade taught}} & {\textbf{Subject taught}} & {\textbf{Location}} \\ \hline
P1 & 6th grade & English Language Arts (ELA) & North Carolina, USA \\ 
P2 & 5th grade & Science & Connecticut, USA \\ 
P3 & 6th-8th grade & Computer Science & Tunisia, North Africa \\ 
P4 & 9th-12th grade & Computer Science & Cuneo, Italy \\ 
P5 & 9th-12th grade & Chemistry and Math & British Columbia, Canada \\ 
P6 & 6th-8th grade & STEM & Florida, USA \\ 
P7 & 6th-8th grade & STEM & Florida, USA \\ 
P8 & - & STEM & Pennsylvania, USA \\ 
P9 & 6th-8th grade & Computer Science & California, USA \\ 
P10 & 9th-12th grade & Career Exploration & Rhode Island, USA \\ 
P11 & 9th-12th grade & Computer Science & Massachusetts, USA \\ 
P12 & 9th-12th grade & Library Science & Rome, Italy \\ 
P13 & 6th grade & History & California, USA \\ 
P14 & 6th-9th grade & Computer Science & Turkey \\ 
P15 & 6th-12th grade & English as a Second Language (ESL) & Pennsylvania, USA \\
\end{tabular}
\end{table}

\subsection{Co-Design Workshop}
The entire co-design workshop spanned two days, Session 1 on the first day and Session 2 on the second day. Session 1 consisted of discussions and a "What is AI" presentation to level set all participants (see Tab. \ref{tab:session-1}), and Session 2 consisted of the co-design activity and an ethics presentation (see Tab. \ref{tab:session-2}). Here, we describe our rationale and the activities in detail.

\subsubsection{Session 1}
Before the first session, we asked participants to complete a pre-workshop questionnaire asking about participants' familiarity with AI, whether they have taught AI in the classroom before, and if so, what their experience was. This was to understand their backgrounds and enable us to tailor the content of Session 1 appropriately. Participants were also given detailed instructions on how to install and use Zoom \cite{zoom}, Slack \cite{slack}, and Miro \cite{miro}---the tools used throughout the entire workshop. We started Session 1 with breaking participants into small groups on Zoom to discuss why participants thought AI is or is not important to teach their students. Having them describe what and why AI was important allowed us to understand their preconceptions about AI and their priorities as teachers. During the ``Let's learn AI'' presentation, participants learned the \emph{Big AI Ideas} \cite{big5}, categories of AI, and how to recognize what is and is not AI. During the ``Let's learn AI tools'' presentation, we demoed four distinct AI learning tools and provided participants with resources and links to explore further. We then used Miro for a card sorting activity \cite{card-sorting}, where we asked participants to generate categories for Google's A to Z of AI cards \cite{atoz}, where categories were limited to subjects taught in the classroom. The card sorting activity showed participants' enthusiasm for integrating AI topics into every classroom subject, including English language arts (ELA), writing and reading, social studies, math, science, economics, and social-emotional learning.

\begin{table}[htb!]
\caption{Schedule for Session 1}\label{tab:session-1}
\begin{tabular}{l|l}
    Time & Activity                          \\ \hline
    15 min  & Introduction                   \\
    20 min  & Why AI? (Discussion)           \\
    50 min  & Let's learn AI! (Presentation) \\
    15 min  & Break                          \\
    25 min  & Card sorting activity          \\
    25 min  & Let's learn AI tools! (Presentation) 
\end{tabular}
\end{table}

\subsubsection{Session 2}
Participants were asked to complete ``pre-work'' before Session 2. Participants had two days to complete their pre-work between Session 1 and Session 2. The pre-work asked participants to explore the rest of the AI learning tools, select one of the tools to go along with a curriculum they currently use or have used in their classrooms, and identify areas where they see potential to teach AI using the selected tool. Participants uploaded their submissions into a shared Google Drive folder. Participants had access to the workshop Google Drive folder, which contained all of the presentations and resources from Session 1, at all times, and could also post questions in the workshop Slack group, which was monitored closely by the researchers. From the pre-work submissions, we selected one idea to develop into an exemplar curriculum (see \citep{gist-appendix}).

For Session 2, participants were split into three groups of 4-5. Each group was asked to analyze the exemplar curriculum and discuss what they noticed. %
The co-design activity part 1 then began with each group responding to a prompt asking them to devise integrated AI curricula for specific subjects. %
We created the prompts from the pre-work submissions and organized the groups such that each would have a domain expert. For example, the group responding to the prompt asking participants to create a curriculum for students who are learning English as a Second Language (ESL) had an ESL teacher, who would be familiar with ESL students' needs. Each group was also paired with a researcher, who provided technical input and answered participants' questions about AI or learning tools. During the ``Ethics \& Diversity'' presentation, we presented definitions of AI ethics, diversity statistics within the field, and resources for teaching and learning AI ethics. Participants then continued working in their groups on their integrated curriculum in co-design activity part 2. 

Every group was successful in producing a first draft of an implementable AI curriculum that integrated with a core subject. The drafts can be found in the appendices \cite{gist-appendix}. Lastly, participants discussed why they thought AI was or was not important to teach for a second time, which acted as a reflection and a way to see if their mindset or preconceptions changed after the workshop. Participants were asked to complete a post-workshop questionnaire that asked how familiar they were with AI, how comfortable they felt teaching AI in their class, as well as feedback on the workshop itself and their demographics (i.e. age, gender, ethnicity).

\begin{table}
\caption{Schedule for Session 2}\label{tab:session-2}
\begin{tabular}{l|l}
    Time & Activity                          \\ \hline
    60 min  & Co-design activity part 1                  \\
    15 min  & Break           \\
    40 min  & Ethics \& Diversity (Presentation) \\
    20 min  & Co-design activity part 2                          \\
    15 min  & Why AI? (Discussion)          \\
\end{tabular}
\end{table}

\subsection{Data Analysis}
Our dataset consists of the audio recordings of the entire co-design workshop, participant questionnaires, and the deliverables of each participant, which include their pre-work submissions and their group work during the co-design activity. All audio recordings were transcribed to text and thematically coded by two researchers using open coding. We specifically examined their process, priorities, and challenges.

Nine out of 15 participants had never taught AI in the classroom. While some participants had had experience teaching AI, they were interested in learning how to allow non-CS students experience AI and integrate AI into their teaching. Participants came into the workshop rating their own familiarity with AI an average of 4.8 out of 7, and finished the workshop with an average rating of 5.8 out of 7. Teachers also rated their confidence about integrating AI into their own curriculum with an average rating of 5.6 out of 7. 

\section{Results and Discussion}
Our teacher participants teach students with diverse needs. The co-design activity prompted rich discussion with three groups completing three curriculum drafts that integrated AI with a topic of their choice. The topics were: (1) ``How Does Data Affect Government Policy?'' (Social Studies Curriculum), (2) ``Learn Vocabulary with an AI'' (Literacy curriculum for students with learning disabilities), and (3) ``Build an AI-powered Pronunciation Application'' (ESL curriculum), as shown in the appendices \citep{gist-appendix}. During the co-designing process, all groups shared certain considerations for the curriculum, though each group addressed them differently. In the first section of results, we answer the first research question by outlining what the shared values and considerations were and showing how each group addressed them. We then answer the second research question by showing how each curriculum effectively integrated AI.

\subsection{RQ\ref{RQ:teaching-needs}: How might we address the values and considerations of K-12 teachers when designing AI curriculum?}

We identified four categories of values and considerations that our teachers had while creating the curriculum drafts: \emph{Evaluation}, \emph{Engagement}, \emph{Logistics}, and \emph{Collaboration}.

\subsubsection{Evaluation} All groups considered student evaluation to be critical to a curriculum. Teachers wanted to see evidence for learning and know their students understand relevant concepts correctly. To do so, teachers first considered their own objectives: ``Do we have an end goal in mind, or like, what do we consider a success?'' (P9). In the ESL curriculum, P12 referred to the \emph{Big AI Ideas} to identify the what the group called the ``AI objective''. P12 and P15 also frequently referred to the exemplar curriculum, suggesting that teachers require frameworks and scaffolding to devise the AI objective. To evaluate students, P5 and P12 both suggested non-traditional forms of evaluation, such as an ``exit interview  or on-the-fly assessments where students talk through all of the details, so we get a really good idea from a conversation with them whether they understand what they were doing'' (P5) and ``an engineer's log where you've got their design and you've got to do it all official'' (P12). In these drafts, teachers wanted to evaluate students on their conceptual knowledge, and not on their technical knowledge.

\subsubsection{Engagement} In a K-12 setting, engagement tends to be particularly challenging, which was a concern for our teachers. P8 and P10 grounded the Social Studies curriculum in law and government discourse by having students review an article around the Crown Act. Introducing context to the project gives students an ``anchor'' (P8) or hook to prompt further inquiry. Other anchors included asking students the ``hard questions'' about real-world applications of AI, such as ``how do Siri and other personal assistants get to be at that point?'' and ``who used the machine learning and designed the app?'' (P7). P5 and P15 both mentioned student-driven learning as a way to leverage students' interests. For example, ``I can see a sixth grader coming in and going, I went to the baseball game and I couldn't say all these words. And they decide they're going to do baseball that day'' (P12). Lastly, multiple groups brought up competition and gamification as effective methods of engagement: ``the class creates a game that students use to quiz themselves on vocab by trying to be better than the system'' and ``module 1 can be a rock paper scissors game so that students get familiar with the interface'' (P2). 

\subsubsection{Logistics}
By logistics, we mean factors that enable the curriculum to be smoothly run in the classroom. Teachers tended to think about how the lesson itself would take shape before addressing which core standards the lesson intended to cover. For example, at the beginning of the co-design, P10 explained that what would be most beneficial was ``thinking of how to structure the lesson and what resources we can use to pull in to have the engagement component''. Most teachers struggled with identifying which technology resources and learning tools to use, for example, whether to use \emph{Machine Learning for Kids} \citep{ml4kids} or \emph{Google Quick Draw} \citep{quick-draw}. Our teacher participants generally looked to the researcher for guidance, suggesting that tools can be more explicit about when and how they can be applied in K-12 classrooms. Teachers also paid close attention to grade-level considerations. They felt more comfortable having older students drive their own learning, but recognized even younger students are capable of deep reflection: ``posing some challenging questions will vary a little depending on age, but you can get pretty deep with some---even fifth graders. They can get into this, and I think it's a good way of opening the door'' (P15).

\subsubsection{Collaboration}
All groups discussed the value of collaboration. In the ESL curriculum, teachers had their students collect data in groups and input the data into multiple models using Machine Learning for Kids. In the literacy curriculum, teachers had every student contribute 10 images to a class dataset to input into Google Teachable Machine. The presence of group work not only helps overcome the need to create many training examples for a machine learning model, but also provides students with opportunities to discuss design and ethics decisions with their peers and teacher. This also aligns with Long and Magerko's design consideration for \emph{Social Interaction} \cite{AILiteracy}.

Teachers also consider how collaboration can be implemented most effectively when designing curricula. For example, P8 described how ``it's important to think about the group size because you want to make sure that students have a voice in the work. And when you start doing large group things those kids that process information internally never get to be heard.'' She went on to describe how, in her experience, ``duos [of students] work really, really well'' and how it is generally better to ``go with smaller groups [of students in the classroom], but if you're using technology [...] you're bound by what you have.''
Thus, it is important to consider how AI tools can best facilitate group work to ensure all students have a chance to contribute and learn.

\subsection{RQ\ref{RQ:integrated-curriculum}: How might AI curricula support teachers when they teach AI?}
During the co-design, teachers made connections between the core subject material (e.g., social studies) and AI in three main ways: \textbf{(1) relating an AI tool or concept to the core subject}, \textbf{(2) relating content from the core subject to AI}, and \textbf{(3) noticing overlapping concepts in AI and the core subject}. For example, P14 related the AI tool, \emph{Arbitrary Style Transfer} \cite{arbitrary-style-transfer}, to the core subject of history when he said, ``If we give an image as input and try to modify [it] according to the old art [using] Style [Transfer]%
[...] This can give us an idea about the history when we look at the picture, [...] but if you change the picture, the students may understand how people thought in the past''. Other teachers related real-life applications of AI to core subjects, like how \emph{YouTube} suggestion algorithms can be ``tunnel visioned'' in what they suggest, similar to how people can be ``tunnel visioned'' when considering politics  or how recidivism risk analysis algorithms \cite{ml-recidivism-richard} can be related to social studies concepts (P10).

Teachers also often made connections by starting with a core subject concept and relating it to AI. For instance, one teacher connected physics data from one of their student's 3D printing projects to an AI flight prediction algorithm (P12). The same teacher also started with an English unit and asked, ``What tools do we know that we [can] connect to language?'', ultimately connecting English to a Shakespeare natural language processing algorithm. Another teacher began with the ELA concept of ``argumentation'' and connected it to the reflection and ``data analysis'' processes in AI (P8).

In terms of overlapping concepts between AI and core subjects, teachers often found connections using the \emph{Big AI Ideas} \citep{big5}. For instance, the Big AI Ideas of \emph{Societal Implications} and \emph{Representation and Reasoning} are also core concepts in social studies. The AI concept of iterative development in ML was also directly connected to the social studies concept of iterative opinion making %
through ``go[ing] back and forth'' (P8) and adjusting beliefs. %

Using these methods of connection, participants co-designed integrated curricula containing AI concepts and supports for teaching core subject requirements. The curricula contained three main points of integration: \textbf{(1) data}, \textbf{(2) reflection}, and \textbf{(3) ethics}.

\subsubsection{Data}
Educational activities often produce data, and AI systems often require data. This provides an obvious access point for AI systems to be integrated into ready-made educational activities. In our co-design workshops, participants used this fact to generate integrated curriculum. %
For instance, in the exemplar curriculum (see Fig. \ref{fig:teaser} and \cite{gist-appendix}) (which was based on a teacher's idea during the workshops) students would produce data as they construct airplanes for a physics activity. The paper airplane dimensions and time-of-flight data would then be used to train a ML model to predict the effectiveness of other potential paper airplanes, %
combining AI systems and physics concepts into a single curriculum.

For the ESL integrated curriculum, students produced data as they were practicing word pronunciation, which was then utilized in a pronunciation teaching app. For example, students would create data by recording saying a word correctly (as guided by a teacher) and incorrectly, which would then train a classification model for an app developed in \emph{MIT App Inventor} \citep{ai-appinv}. This app would then be used to further help students learn correct pronunciation. Future AI-integrated curricula might consider utilizing the data inherent in core curricula activities, such as speech pronunciation data, to teach AI. This may be through teaching data-related AI competencies, like \emph{Data Literacy}, \emph{Learning from Data}, and \emph{Critically Interpreting Data} \citep{AILiteracy}, or through using data to train ML models, which can teach other AI competencies.

For the literacy curriculum for students with learning disabilities, participants also used data from core curriculum---vocabulary words---to integrate AI concepts. From the vocabulary words, students would find relevant images, %
generating further data, and use this to train a classification model. %
This addressed the aforementioned data-related AI competencies, as well as other competencies, including the \emph{ML Steps} and \emph{Human Role in AI}, in addition to relevant English literacy concepts.

\subsubsection{Reflection}
Another point of AI integration was student reflection on core curriculum content and AI methods. Many common core standards as well as AI competencies can be addressed through student reflection. For example, the common core standard, 1-ESS1-1: ``Use observations of the sun, moon, and stars to describe patterns that can be predicted.'' \citep{ngss-nsta}, and the AI concept, ``Learning from Data'', could be addressed by reflecting on patterns in a constellation classification model's input and output. In the exemplar curricula, students were asked to reflect on what did and did not work and why, and on the real-world implications of a biased dataset in airplane development. This reflection addressed both a standard from the common core, 3-5-ETS1-3: ``Plans and carries out fair tests in which variables are controlled and failure points are considered to identify aspects of a model or prototype that can be improved'' as well as a number of AI literacy competencies, including \emph{AI Strengths \& Weaknesses}, \emph{Critically Interpreting Data}, and \emph{Ethics} \citep{AILiteracy}.

Teachers also used this method to integrate AI concepts into the social studies curriculum. For example, students were asked to reflect on the amount of data in each image category, social norms and peer opinion, people's ability to access resources, and consensus agreement in this curriculum. These reflection questions address a number of the AI competencies, including \emph{Data Literacy}, \emph{Critically Interpreting Data}, and \emph{Ethics} \citep{AILiteracy}, as well as core social studies and English language arts standards, including NSS-EC.5-8.1: \emph{Scarcity}, NSS-C.5-8.3: \emph{Principles of Democracy}, NL-ENG.K-12.4: \emph{Communication Skills}, and 
NL-ENG.K-12.7: \emph{Evaluating Data} \citep{education-world-standards}.

\subsubsection{Ethics}
The final point of integration we present is through ethics, which is one of the AI literacy competencies \citep{AILiteracy}. Ethics can also be found in many common core standards \citep{ngss-nsta,education-world-standards}. For example, environmental ethics can be found in life science standards (K-ESS3-3: ``Communicate solutions that will reduce the impact of humans on the land, water, air, and/or other living things in the local environment.''), %
and engineering standards (MS-ETS1-1: ``Define the criteria and constraints of a design problem [...] taking into account [...] potential impacts on people and the natural environment'') \citep{ngss-nsta}. Furthermore, social justice principles, which are highly related to AI ethics, are commonly advocated for within standards-based K-12 education \citep{social-justice-standards-based-dover,social-justice-advocacy-dixon}. 
By teaching ethical principles with respect to AI, teachers can also address standards related to the common core.

Each curricula designed in the workshops had an ethics component. %
In the exemplar, students would engage in a brainstorming session about how AI bias affected the accuracy of ML models and relevant implications in the real world. Similarly, the ESL curriculum addressed ethics through discussing AI bias, socioeconomic norms for ``correct'' pronunciations, and the implications of an AI system judging people's pronunciations in the real-world. The social studies curriculum was developed around the ethics of the ``CROWN Act'' \citep{crown-act-edwards}, what it means for students to design AI algorithms to classify outfits and hairstyles as ``professional'' or ``unprofessional'', and how this might affect different people groups. The literacy curriculum for students with learning disabilities addressed ethics through discussion about the accuracy of the image classification system and reasons for any bias observed. Each of these curricula touched on environmental, social justice or other ethical issues, addressing both AI and common core ethics standards.

From the workshops, we found that teachers were highly interested in teaching ethics (e.g., the social studies curricula was entirely focused on ethics); however, they also seemed apprehensive about actually implementing ethics activities in the classroom. For example, P5 described how there is a ``barrier that comes up for teachers'' when `` kids often bring ethics up with questions and sometimes teachers will avoid it because they’re afraid to say something wrong [...] even though those discussions would be so rich.'' Nevertheless, P5 also mentioned how if it was in a ``planned lesson'', it would be ``less scary because you know what you’re going to say''. Designing scaffolding for AI ethics lessons would not only enable core curricula integration, but would also empower teachers to more confidently teach students about ethics.

\subsection{Reflecting on Remote Co-design}
Due to Covid-19, we organized and ran this co-design workshop completely remotely. Among our activities, the perceived helpfulness from most to least helpful was: Presentations (11 votes), Co-design activity (9 votes), Why AI? and Ethics discussions (both 7 votes), and the Card sorting activity (4 votes). Participants also indicated meeting like-minded educators from around the world and having access to the list of tools and links to be particularly rewarding takeaways. Overall, we noticed a slight increase in familiarity with AI after the workshop and a high level of confidence for integrating AI into their classrooms, though we did not establish that baseline. When asked if the workshop changed their opinion about teaching AI, teachers cited ``introducing AI is the gateway to so much learning...now I am seeing and starting to understand the vast world of opportunities that exist for coding beyond being video game designers'' (anonymous), as well as seeing the necessity of teaching AI and understanding that AI can be accessible to not ``just the computery people'' (anonymous). 

At the beginning of the workshop, we established norms as an entire group to make facilitating easier. For example, setting expectations for ``warm'' calling to ensure equal representation of voices in the room meant participants expected to be called on to share their thoughts. Other norms included being present, having discussions in breakout rooms, and keeping cameras on. Since most teachers were unfamiliar with teaching AI, we grouped them into smaller groups of 4-5 so that each group could have a researcher co-designing with them. However, this meant some teachers worked on curriculum that was unrelated to their discipline. We believed this trade off was necessary given the complexity of the task, and was mitigated by the benefits of collaboration. This setup may have worked better if teachers from the same school joined, and groups could be formed by school. Several teachers also requested more time to play with the AI learning tools and digest the presentations. This could have been addressed by scheduling more time between Session 1 and 2, so teachers would have more time to complete the pre-work for Session 2. One suggestion from a participant was to introduce the AI tools using a jigsaw game where every teacher explores an assigned AI tool and presents it back to the group. 

\subsection{Broader Implications and Limitations}
The above findings contribute to the under-explored need to collaborate with teachers when designing AI curriculum, as well as the potential for AI to be integrated into K-12 core curriculum. Combining teaching expertise with research expertise through co-design allowed for thinking beyond the context of a research study and into actual classrooms. The adoption of learning tools and AI curriculum is influenced by complex factors outside the locus of control of the people creating the tools (i.e. designers and researchers) and the people using the tools (i.e. teachers). However, without teacher buy-in, adoption in the classroom would be impossible, and understanding their contexts is necessary and often understated. Through this co-design, teachers experienced the potential for AI to be embedded in subjects like social studies and English, which could allow non-CS and non-technical learners to experience AI in new classrooms. This method also leverages students' existing interests in non-technical subjects as a pathway into AI. This work serves as a push for further explorations to expose a wider range of students to AI.

While the above findings could provide useful insights to AI education researchers and designers, we acknowledge the limitations of our study. Because our explorations focused on integrating AI with non-technical subjects with a small group of teachers, applying and extending general implications beyond this context should be done with caution. 

\section{Conclusion}
In this paper, we explored the needs of teachers in K-12 classrooms and how AI education can be integrated with existing core curriculum. We engaged K-12 teachers and researchers in a two-day co-design workshop, where we co-created lesson plans that embedded AI concepts into curricula for social studies, ESL, and literacy for students with learning disabilities. We found that teachers value curriculum that address evaluation and engagement of students, which could be built into the learning tool or curriculum. Teachers also successfully connected AI with their subject by having students examine subject-related datasets, as well as reflect on real-world implications and AI ethics. Our work highlights an opportunity to increase accessibility of K-12 AI education by embedding AI into core subjects (e.g., English, social studies), and reaching students outside of CS and technology classrooms.

\begin{acks}
We thank the teachers who were a part of this study; Randi Williams, who provided co-design guidance; and Hal Abelson, who made this work possible. %
\end{acks}

\bibliographystyle{ACM-Reference-Format}
\bibliography{biblio}

\end{document}